\documentclass[twocolumn]{article}

\usepackage[margin=0.75in]{geometry}

\usepackage{amsmath, amssymb}
\usepackage{graphicx}
\usepackage{booktabs}
\usepackage{xcolor}
\usepackage{caption}
\usepackage{array}
\usepackage{fancyvrb}
\usepackage{framed}
\usepackage{multicol}
\usepackage{float}
\usepackage{verbatim}
\usepackage{tabularx} 

\usepackage{listingsutf8}
\usepackage{listings}
\newcommand{\keywords}[1]{\vspace{2mm}\noindent\textbf{Keywords:} #1\par}

\usepackage[backend=bibtex, style=numeric, sorting=none]{biblatex}
\title{CCCI: Code Completion with Contextual Information for Complex Data Transfer Tasks Using Large Language Models}

\author{
Hangzhan Jin\thanks{hangzhan.jin@polymtl.ca} \and Mohammad Hamdaqa\thanks{mhamdaqa@polymtl.ca}\\
PolyTechnique Montréal, Montréal, Canada
}

\begin{document}

\maketitle

\begin{abstract}
Unlike code generation, which involves creating code from scratch, code completion focuses on integrating new lines or blocks of code into an existing codebase. This process requires a deep understanding of the surrounding context, such as variable scope, object models, API calls, and database relations, to produce accurate results. These complex contextual dependencies make code completion a particularly challenging problem. Current models and approaches often fail to effectively incorporate such context, leading to inaccurate completions with low acceptance rates (around 30\%). For tasks like data transfer, which rely heavily on specific relationships and data structures, acceptance rates drop even further. This study introduces CCCI, a novel method for generating context-aware code completions specifically designed to address data transfer tasks. By integrating contextual information, such as database table relationships, object models, and library details into Large Language Models (LLMs), CCCI improves the accuracy of code completions. We evaluate CCCI using 289 Java snippets, extracted from over 819 operational scripts in an industrial setting. The results demonstrate that CCCI achieved a 49.1\% Build Pass rate and a 41.0\% CodeBLEU score, comparable to state-of-the-art methods that often struggle with complex task completion.
\end{abstract}



\maketitle


\section{Introduction}
Recent advancements in Large Language Models (LLMs) like GPT and Copilot \cite{github_copilot_2021} have demonstrated the potential in code completion tasks \cite{liu_neural_2020, abedu_llm-based_2024}, thereby enhancing developer productivity. Despite these advancements, the acceptance rates for these tools remain low \cite{mozannar_when_2024, liang_large-scale_2024}, primarily due to their generation of generic, context-agnostic code. Prior studies in automated code completion face several significant limitations. Many approaches rely solely on integrating existing libraries \cite{tang_domain_2023, liu_codegen4libs_2023}, recommending APIs  \cite{ma_compositional_2024, jain_mitigating_2024} based on those libraries without considering data structure or relations. Others require extensive labelled data for fine-tuning \cite{van_dam_investigating_2024} or pre-training \cite{jiang_treebert_2021, wang_ast_2021, liu_language_2020} and demand developers to provide pseudo-code \cite{yu_codereval_2024, jackson_creativity_2024}, which is labor-intensive and not scalable for large systems. Additionally, some methods focus only on retrieving \cite{lu_reacc_2022} or searching \cite{ryan_code-aware_2024} similar code examples for code suggestions, and some methods pay more attention to predicting code completion invocation \cite{van_dam_investigating_2024}, neglecting the contextual information like code comments \cite{obrien_are_2024, ahmed_automatic_2024} embedded in the applications of enterprises. These methods fail to leverage the real-time data structure and relations present in current software development, resulting in code completions that often lack relevance, especially in data transfer tasks. This gap highlights the need for a more sophisticated approach that can integrate detailed contextual information to enhance the relevance and applicability of code completion.

Given the aforementioned challenges, our study aims to refine the integration of contextual information into LLMs. We hypothesize that this enriched context can improve the utility and accuracy of automatic code completion in data transfer tasks. By addressing these research gaps, our work seeks to pave the way for more sophisticated, context-aware tools that can genuinely enhance developer efficiency. The significance of our research lies in addressing this gap by integrating contextual information—such as database table relations, existing object models, and specific API usage—directly into LLMs. This approach aims to produce higher accuracy code, thereby potentially increasing the practical utility of these models in real-world software development environments. The main contributions of our paper are summarized as follows:

\begin{itemize}
\item We propose a retrieval-augmented code completion approach that can retrieve code from the current project and its complex dependencies and integrate different LLMs to improve code completion performance.
\item We perform a comprehensive evaluation of our approach, and the results illustrate that our approach improves the GPT-4o by up to 142.6\% in terms of CodeBLEU, the Build Pass rate improved dramatically from 0\% to 49.1\%.
\item We examine the performance of our approach in six popular open-source and closed-source LLMs, and the results show our approach can produce a noticeable improvement in those models compared to the GPT-4o model with an original prompt.
\end{itemize}

The structure of this paper is organized as follows: Section II introduces the challenges of code completion for data transfer tasks necessary for understanding the concepts discussed. Section III details our research methodology. Section IV presents the empirical evaluation of the CCCI method. Section V reviews related work to contextualize our study within the current state of research. Section VI identifies potential threats to the validity of our study, and Section VII concludes the paper with a summary of our contributions and future research directions.

\section{Challenges of Code Completion for Data Transfer Tasks}

\begin{figure}[h]
\centering
\includegraphics[width=\linewidth]{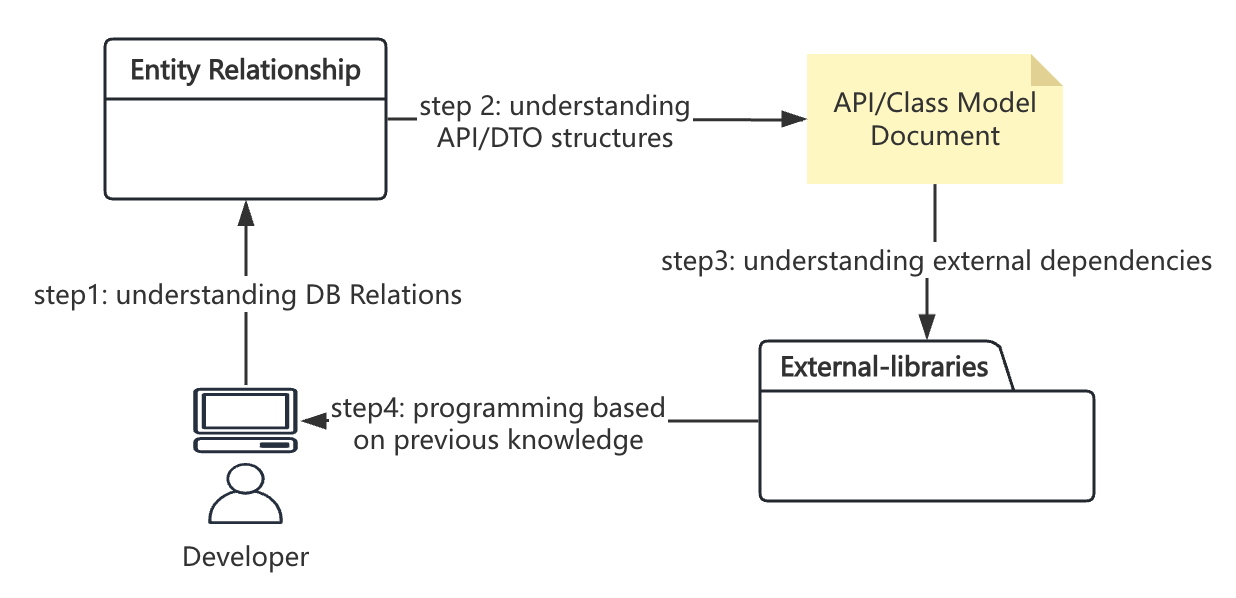}
\caption{A common data transfer scenario}
\label{fig:complex_process}
\end{figure}

To illustrate the challenges of code completion for data transfer tasks, we will start by explaining a case study of a Warehouse Management System (WMS) and use this case study to highlight the challenges as well as showcase the steps in our approach. Data transfer tasks are repetitive and complex tasks that developers frequently encounter in real software development, transforming one or more input objects from different data sources like APIs or databases into an output object by mapping fields to the required format \cite{ngu_semantic_2010, lorenzo_data_2009}. This involves converting front-end request parameters into the format required by third-party services and modifying service outputs to meet the website's needs. As shown in Figure \ref{fig:complex_process}, a developer needs to understand the database Relations based on Entity Relationship Graph, API information with data structures in both document and source code, and external libraries to guide programming for the tasks. This process demands extensive time to understand the contextual information \cite{bogomolov_long_2024, liu_context_2023} for programming. While capable of generating high-quality code, it often requires developers to invest substantial effort in preparing detailed prompts and contextual background, which can be as time-consuming as manual coding, and the quality of generated code depends on developers' skill to write prompts \cite{min_beyond_2024, coignion_performance_2024, grewal_analyzing_2024}, making it difficult for them to adopt code completion to finish data transfer tasks manually. 

\subsection{Case Study}\label{case-study}

We explore the application of our CCCI methodology within a real enterprise's WMS. A WMS typically integrates with numerous external systems and is designed to optimize warehouse operations by streamlining tasks such as inventory management, order fulfillment, and logistics. The WMS under study in our research includes 111 tables, 2204 columns, and nine external libraries. In such a system, using LLMs for data transfer tasks needs to analyze the hierarchical data structures and flatten them into text by recursively retrieving data file information within current projects and the nine third-party dependencies, then combine the text with DB relationships and task definitions to construct the intricate prompt to guide LLMs implementing the code completion. The study's goal is to automate this process and use the WMS system to verify our approach. 

We use a data transfer task that maps the classes and field information from four input models: InventoryInfoDTO, SKUInfoDTO, UserDTO, and WarehouseDTO~(Listing 1) to the output model InventoryResponseDTO (Listing 2) as a showcase to illustrate the challenges of this task due to complex data structures.

\textbf{Listing 1: An example of input models}
\begin{framed}
\begin{Verbatim}[fontsize=\small]
class InventoryInfoDTO {
    String inventoryName;
    int availableQuantity; }
class SKUInfoDTO {
    int inventoryId;
    String skuName;
    int ownerUserId;
    String ownName; }
class UserDTO {
    String name;
    String contactInfo; }
class WarehouseDTO {
    int inventoryId;
    String warehouseLocation;
    String managerName; }
\end{Verbatim}
\end{framed}

\textbf{Listing 2: An example of an output model}
\begin{framed}
\begin{Verbatim}[fontsize=\small]
class InventoryResponseDTO {
    String name;
    int availableQuantity;
    SKUInfo sku;
    WarehouseInfo warehouse; }
\end{Verbatim}
\end{framed}

\begin{figure}[h]
\centering
\includegraphics[width=.6\linewidth]{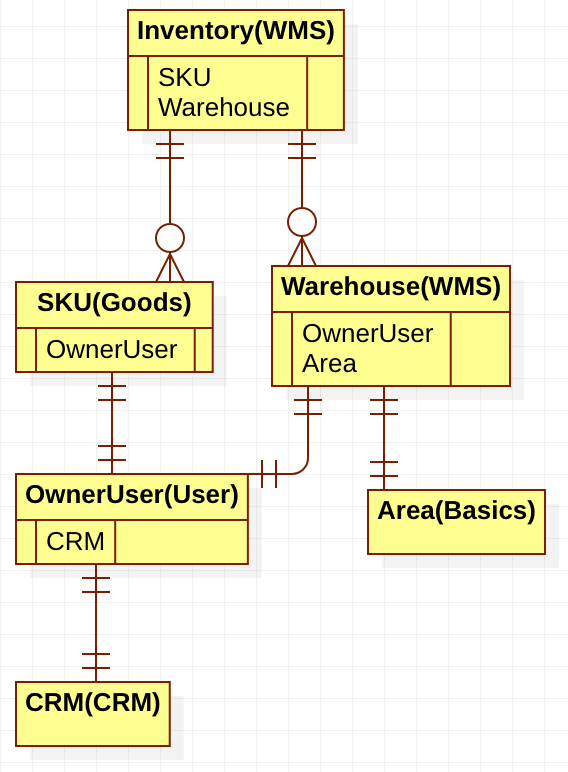}
\caption{An example of complex data structures}
\label{fig:complexrelations}
\end{figure}

\subsection{Challenges}\label{Challenges}

As shown in Figure \ref{fig:complexrelations}, the Warehouse Management System (WMS) involves a complex relationship among several interconnected objects. The Inventory object includes both SKU (Stock Keeping Unit) and Warehouse objects. The SKU, categorized under the Goods module, and the Warehouse, located within the WMS service, both incorporate an OwnerUser as a sub-object. Additionally, the Warehouse object has an Area object, categorized under the Basics module, as a sub-object. Moreover, the OwnerUser object contains CRM (Customer Relationship Management) details housed in the CRM module. To transfer many input objects from different systems to such an intricate output object, there are four difficulties that need to be overcome:

\textbf{• Complicated external dependencies.} Many projects involve external libraries or third-party dependencies distributed in many systems like Customer Relationship Management(CRM), and current tools often fail to accurately account for these, resulting in incorrect or incomplete code \cite{zhang_codefort_2024}.

\textbf{• Complex data structure.} AI models struggle to accurately generate code  \cite{lian_uncovering_2024, ding_semcoder_2024}, especially for projects with complex, hierarchical data structures and interdependent objects(e.g., the Inventory Data Object Model~(DTO) includes multi-layers of DTOs), these models are scattered across different projects; leading to errors or inaccurate code.

\textbf{• Lack of Database Relations.} Code generation tools may overlook or misinterpret database relationships, which are crucial for tasks like data mapping and object-relational mapping (ORM).

\textbf{• Complexity of Prompt Formulation.} Prompt Formulation is a complex and ad-hoc task because prompt formulation requires manually providing project-specific details that the LLM lacks, making each prompt more complex. Crafting detailed prompts demands time, domain knowledge, and iterative refinement to ensure precision. Effective prompts guide the LLM away from common incorrect assumptions, requiring deep expertise in both prompt formulation and model behavior. Prompt designers must adjust and test multiple times to achieve accurate, complete results. Creating prompts that adapt to changing project structures or live data integrations is technically challenging. 

The study's goal is to automate this process and use the WMS system to verify our approach.

\begin{figure*}[htbp] 
\centering
\includegraphics[width=\textwidth]{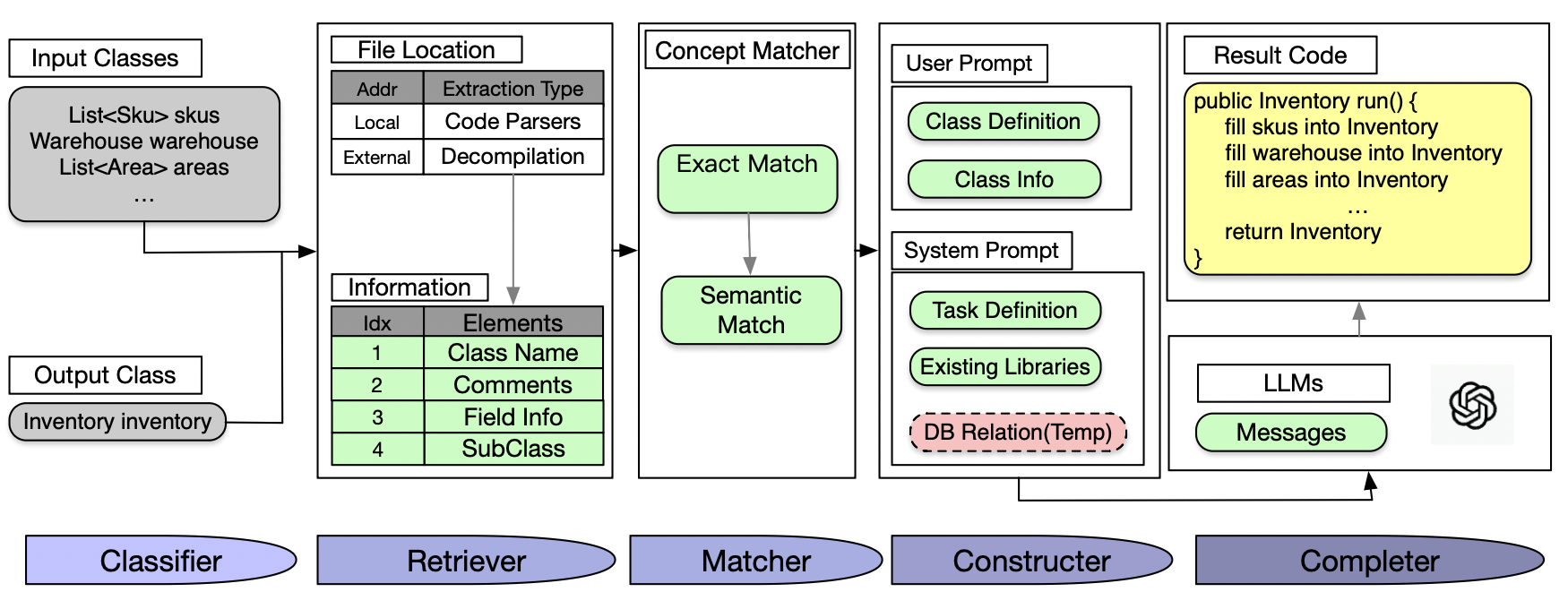}
\captionsetup{justification=centering}
\caption{Overview of the CCCI process}
\label{fig:ccci-framework}
\end{figure*}

\section{Methodology}\label{methodology}

As we discussed in the previous section, the complexities of data structure, external dependencies, prompt formulation and the lack of DB relations are obstacles to the LLMs undertaking code completion for data transfer tasks. 


In this section, we describe our research methodology called CCCI (Code Completion with Contextual Information), which is designed to enhance code completion by integrating contextual information into Large Language Models (LLMs). 

CCCI addresses these issues by automating the inclusion of contextual information, thereby enhancing the relevance and accuracy of the generated code. CCCI takes as input the \textit{project link}, which provides access to the project's source files, dependencies, and database tables, as well as the 
\textit{Task Definition}, which is a high-level description of what needs to be achieved, as shown in Listing 3. 

\textbf{Listing 3: The task definition}
\lstset{language=[LaTeX]TeX}
\begin{lstlisting}
Task Overview:
Given the following project <project link>
Generate Java code to transform Input DTOs into Output DTO.

Input/Output Description:
- Input: Multiple DTO objects or lists (e.g., Listing 1).
- Output: Transformed DTO object (e.g., Listing 2).

Additional Context (e.g., "Use BeanUtils.copyProperties for identical fields"). 
\end{lstlisting}


Figure \ref{fig:ccci-framework} illustrates the CCCI approach.  CCCI consists of five components that streamline context-aware code completion. It begins with the (i) \textbf{File Classifier}, which identifies and categorizes data sources into project files or external libraries. The (ii) \textbf{Information Retriever} extracts hierarchical data structures, including class details, field types, and relationships, from the categorized data sources. For project files this can be through parsing, and from external libraries by first decompilation then extraction. Next, the (iii) \textbf{Concept Matcher} aligns input and output DTOs fields through exact and semantic matching to establish detailed mappings. These mappings are transformed into structured, text-based prompts by the (iv) \textbf{Prompt Constructor}, combining hierarchical relationships and predefined rules to guide the LLM. Finally, the (v) \textbf{Code Completer} generates contextually relevant code using the enriched prompt, addressing challenges in complex data transfer tasks. The rest of this section explains each of these components in detail.



\subsection{Classifier}\label{Classifier}
The first component in CCCI is Classifier, which is to overcome the challenge of complicated and distributed dependencies across different projects by classifying and locating the DTO source files and then appropriate techniques to be applied to extract hierarchical data structures from both current projects and third-party libraries later. Classification determines whether the inputs and output DTO Source Files belong to the current project or external dependencies. The process begins by recursively searching the DTO files in the current project. If a file is found within the WMS system, it is tagged as a Local File, indicating its direct usage within the project's scope. Conversely, files not found are tagged as External Dependencies. Consider a Warehouse Management System (WMS) project with the following DTOs:
\begin{itemize}
\item {\texttt{Input DTOs}}: InventoryInfoDTO, SKUInfoDTO, UserDTO, WarehouseDTO.
\item {\texttt{Output DTO}}: InventoryResponseDTO.
\end{itemize}
The Classifier begins by searching the project directory for each DTO. For example:
\begin{itemize}
\item InventoryInfoDTO and SKUInfoDTO are found in the local project directory and are classified as Local.
\item UserDTO and WarehouseDTO are found in external libraries user-api.jar and warehouse-api.jar respectively, and are classified as External.
\end{itemize}
The output of the Classifier is a detailed mapping: 
\begin{framed}
\begin{verbatim}
InventoryInfoDTO: Local
InventoryResponseDTO: Local
SKUInfoDTO: External (goods-api.jar)
UserDTO: External (user-api.jar)
WarehouseDTO: External (warehouse-api.jar)
\end{verbatim}
\end{framed}
This classification ensures subsequent Retriever can extract the class information from the appropriate sources.

\subsection{Retriever}\label{Retriever}

The goal of the second phase is to retrieve complicated data structures from the current project and its dependencies; we employ the tagged files preprocessed by Classifier to Retrieve data structures from the root node to children nodes and their children nodes recursively. These nodes include class names, field types, field names, and comments in each layer, which enrich the contextual information of the input and output DTOs (as demonstrated in Listing~4) utilized for the next field matching. We use different strategies to handle local and external files. For local files, we employ JavaParser as a code parser to extract detailed information from these files, facilitating a thorough understanding of the project's internal structure. However, files distributed within third-party libraries pose a unique challenge as they are complicated, compiled and lack accessible source text. To address these problems, we employ decompilation techniques to analyze these compiled files. In this context, the Java Reflection Mechanism is implemented to decompile and extract hierarchical data structures from these files, such as annotations, fields, and their types. This dual approach ensures a comprehensive data structure extraction from both in-project (the WMS system) source files and external compiled libraries (third-party dependencies). 

\textbf{Code parsers} are pivotal in our methodology for analyzing source code from current projects. They textually parse code to extract structural and semantic details such as class names, comments, field names, and field types. This process provides a rich context that enhances our model’s ability to generate accurate and contextually relevant prompts. Serving as a fundamental component for data preparation, code parsers accommodate a variety of programming languages and environments.

\textbf{Decompilation Techniques} are employed to handle compiled DTO files, especially those from external libraries that typically lack source comments. These techniques enable us to decompile the compiled files to access their underlying structure, extracting metadata such as class names, field names, field types, and annotations. This approach helps recover structural information provided by annotations, compensating for the absence of direct commentary and ensuring a thorough extraction of class information critical for our processing needs.

Example: for the local DTOs identified in the Classifier, we have local DTO - InventoryInfoDTO:
\begin{framed}
\begin{verbatim}
//The inventory information
class InventoryInfoDTO {
    // Name of the inventory
    String inventoryName; 
    // Stock available
    int availableQuantity; 
}
\end{verbatim}
\end{framed}
The Retriever extracts:
\begin{framed}
\begin{verbatim}
Class: InventoryInfoDTO
Fields:
  - inventoryName: String
  - availableQuantity: int
Comments:
  - inventoryName: "Name of the inventory"
  - availableQuantity: "Stock available"
\end{verbatim}
\end{framed}
External DTO - UserDTO (from user-api.jar): after decompiling by Java Reflection, the Retriever extracts:
\begin{framed}
\begin{verbatim}
Class: UserDTO
Fields:
  - name: String
  - contactInfo: String
\end{verbatim}
\end{framed}
This information is stored in a structured format like Listing 4, ready for the next component.

\subsection{Matcher}\label{Matcher}

The Matcher facilitates accurate data transfer between input and output Data Transfer DTOs; our approach uses a two-step field-matching process:
\textbf{Step 1: Exact Field Matching}
This initial step automatically transfers fields with identical names between the input and output DTOs, ensuring straightforward and accurate data alignment when field names match exactly. 
\textbf{Step 2: Semantic Field Matching}
In cases, as defined in Listing 3 for the second rule, where field names differ but represent the same concept, a semantic matching approach leverages large language models (LLMs) to predict and align fields based on contextual clues, such as class information or field annotations. This step enhances flexibility by enabling matches even when fields have different names but similar meanings, reducing the need for manual mapping adjustments.
The Semantic Field Matching is designed to identify initial correspondences between source and target Data Transfer Objects, providing a foundation for creating compatible structures. This matching process involves four key steps: \textbf{1.	Selecting representative concepts by filtering redundant fields:} This step focuses on selecting only the relevant class information directly related to the input and output parameters while eliminating redundancy. For example, if multiple DTOs reference the same class(e.g., OwnerUser in Figure \ref{fig:complexrelations}), the class is selected only once. Similarly, if a superclass is already included, its subclasses are excluded unless they add distinct semantic meaning. This ensures the representation is both concise and meaningful. \textbf{2.Generating definitions that include annotations, field names, and comments:} Class information such as fields, comments, and hierarchical context is retrieved after identifying the representative concepts. This information, provided by the previous component in the pipeline, forms the basis for understanding the semantics of each concept. \textbf{3.Computing embeddings and cosine similarity scores for semantic alignment:} The selected concepts and their associated definitions are combined into structured representations. These representations are then converted into dense vector embeddings, capturing their semantic meanings. A cosine similarity \cite{elhamlaoui_model_2021} measure is applied to compare the vectors, calculating how closely aligned the source and target representations are. \textbf{4.Ranking and selecting the most relevant matches:} The correspondences with the highest similarity scores are identified and ranked. These top-ranked correspondences are then selected and incorporated into the context for further refinement. Focusing on the most relevant matches, this step ensures that subsequent processing can combine these correspondences into instructions, which guides LLMs to match the semantic information to generate code from different DTOs.

Example: For the input DTOs InventoryInfoDTO and SKUInfoDTO and the output DTO InventoryResponseDTO:
Exact matching matches the fields with the exact same name:
\begin{lstlisting}
InventoryInfoDTO.warehouseName →
InventoryResponseDTO.warehouseName
InventoryInfoDTO.availableQuantity → 
InventoryResponseDTO.availableQuantity
\end{lstlisting}
Semantic matching matches the fields with similar meanings even if their names are different:
\begin{lstlisting}
InventoryInfoDTO.inventoryName → 
InventoryResponseDTO.name
SKUInfoDTO.skuName → 
InventoryResponseDTO.sku.skuName
SKUInfoDTO.user.name → 
InventoryResponseDTO.sku.ownName
\end{lstlisting}

The resulting mapping table merges all the fields matched, including both the exact matching and semantic matching:
\begin{lstlisting}
Input Field                        → Output Field
InventoryInfoDTO.warehouseName     → InventoryResponseDTO.warehouseName
InventoryInfoDTO.inventoryName     → InventoryResponseDTO.name
InventoryInfoDTO.availableQuantity → InventoryResponseDTO.availableQuantity
SKUInfoDTO.skuName                 → InventoryResponseDTO.sku.skuName
SKUInfoDTO.user.name               → InventoryResponseDTO.sku.ownName
\end{lstlisting}

\subsection{Constructor}\label{Constructor}

The Constructor formulates a structured prompt that combines the task definition and rules into a system prompt, as well as mappings and contextual information generated by Matcher into the user prompt. This prompt is used as detailed instruction to guide LLMs in generating contextually relevant code. The Constructor organizes the mappings into a hierarchical data model, which is then converted into a text-based prompt. This prompt includes task-specific instructions. For example, to alleviate the issue of token limitation, we use BeanUtils.copyProperties for exact matches. It also incorporates rules to avoid redundant operations or unnecessary class declarations.
Example: Using the mapping table from the Matcher, the Constructor generates the following prompt, the input DTOs are the input for the matching task; the output DTO is the output that the matching task is supposed to return; the rules are to guide LLMs completing the both exact matching and semantic matching. 
\begin{lstlisting}
Task: Map input DTOs to the output DTO.
Input DTOs:
  - InventoryInfoDTO: [warehouseName, inventoryName, availableQuantity]
  - SKUInfoDTO: [skuName]
  - UserDTO: [name]
Output DTO:
  - InventoryResponseDTO:
      - warehouseName → InventoryInfoDTO.warehouseName
      - name → InventoryInfoDTO.inventoryName
      - availableQuantity → InventoryInfoDTO.availableQuantity
      - sku.skuName → SKUInfoDTO.skuName
      - sku.ownName → SKUInfoDTO.user.name
Rules:
  1. Use BeanUtils.copyProperties for fields with identical names.
  2. Manually map fields with different names but similar semantics.
\end{lstlisting}

\subsection{Completer}\label{Completer}
The Completer employs the structured prompt formulated by Constructor to produce the final code. This step leverages the prompt with detailed mappings and rules to guide LLMs in generating syntactically and contextually correct code. The prompt is fed into LLMs, which interpret the instructions and generate code that adheres to the rules and mappings.
Example: Given the prompt, the LLM generates:
\begin{lstlisting}
InventoryResponseDTO response = new InventoryResponseDTO();
response.setName(inventoryInfoDTO.inventoryName);
// Copy identical fields
BeanUtils.copyProperties(inventoryInfoDTO, response);

// Map nested objects
SKUInfo skuInfo = new SKUInfo();
skuInfo.setSkuName(skuInfoDTO.getSkuName());
skuInfo.setOwnName(user.getName());
response.setSku(skuInfo);
\end{lstlisting}
This final code ensures that all mappings are implemented correctly, including handling of nested objects.

\textbf{Listing 4: The example of class information retrieved by CCCI}
\begin{framed}
\begin{Verbatim}[fontsize=\small]
Input Parameters:
// Query inventory flow table
- private com.xx.ResponseList<com.xx.
InventorySkuItemFlow> inventorySkuItemFlow;    
// Query warehouse areas
- private java.util.List<com.xx.WarehouseArea>
  warehouseAreaList; 
// Query warehouse locations
- private java.util.List<com.xx.
WarehouseLocation> warehouseLocationList;    
// Query warehouse
- private com.xx.Warehouse
  warehouse;    
Output Information:
- private List<com.xx.api_2694416191343104
.resp.Content> contents;

Entity Details:
- Entity:Inventory Info DTO:com.xx.Inventory
  Fields: 
  id:primary key:long, 
  name:name of inventory:String,
  sku: sku info object: com.xx.SkuDTO
- Entity:SKU for goods:com.xx.SkuDTO
  Fields: 
  inventoryName:name of inventory:String,
  ownName:owner name:String
- Entity: com.xx.UserInfoDTO
  Fields: 
  name:username:String
- Entity:Area Info:com.xx.WarehouseArea
  Fields: ...
- Entity: com.xx.WarehouseLocation
- Entity: com.xx.Warehouse
- Entity: com.xx..resp.Content
- Entity: com.xx.SkuInfoVO
- Entity: com.xx.WarehouseAreaVO
\end{Verbatim}
\end{framed}

Listing 4 illustrates the retrieved class information, including data structure of inputs and output DTO(s), the comments and package of DTO(Entity), fields of DTO, comments and type of fields. 

\textbf{Listing 5: An example of database table relations}
\begin{framed}
\begin{verbatim}
Warehouse Domain:
- warehouse (Warehouse)
  |-> warehouse_area (Warehouse Area): 
  1:N relationship
  |-> warehouse_dock (Dock): 
  1:N relationship
- warehouse_area (Warehouse Area)
  |-> warehouse_location (Warehouse 
  Location): 1:N relationship
\end{verbatim}
\end{framed}

\section{Empirical Evaluation}\label{Empirical Evaluation}

In the empirical evaluation section, we adapt the benchmarking study \cite{benchmarking} and utilize a dataset consisting of 819 scripts from the WMS systems, which have been developed and deployed. These scripts, with their respective input and output objects, serve as the reference for the generated code and standard structure for our CCCI method. Using this approach, we generate new scripts and then employ CodeBLEU, BLEU-4, Edit Similarity, and Build Pass to score the generated scripts. We use the same 289 scripts from the whole system to generate code. The results will illustrate the accuracy of the generated scripts compared to the original ones and the impact of different models.

\begin{table}[ht]
\centering
\caption{The dataset for WMS system}
\label{tab:dataset for WMS system}
\begin{tabular}{@{}cccc@{}}
\toprule 
\textbf{System} & \textbf{Snippets} & \textbf{Tables} & \textbf{Dependencies} \\ 
\midrule 
WMS & 819 & 111 & 9  \\
\bottomrule 
\end{tabular}
\end{table}

As the table shows, the WMS system we use for the experiment includes 819 reference code snippets and 111 tables, and it depends on nine third-party libraries (dependencies); such an intricate system emphasizes the challenge of retrieving contextual information from both the current project and its third-party dependencies.

\textbf{Listing 6: An example of snippets}
\begin{lstlisting}
xxx.RequestModel param = new xxx.RequestModel();

Integer skuForm = Optional.ofNullable(asnOrder)
        .map(AsnOrder::getAsnOrderItems)
        .filter(CollectionUtils::isNotEmpty)
        .map(it -> it.get(0))
        .map(AsnOrderItem::getSkuForm)
        .orElse(SkuFormEnum.GOOD.getCode());

ShelfOrder shelfOrder = new ShelfOrder();
shelfOrder.setRelatedOrderId(result.getId());
shelfOrder.setRelatedOrderType(RelatedOrderTypeEnum
.INBOUND_ORDER.getCode());
shelfOrder.setShelfItems(requestModel.getSkus()
        .stream()
        .filter(Sku::getIsCheck)
        .map(it -> {
            ShelfItem shelfItem = new ShelfItem();
            shelfItem.setQuantity(it.getQuantity());
            shelfItem.setSku(it.getSku());
            shelfItem.setSkuForm(skuForm);
            return shelfItem;
        })
        .collect(Collectors.toList()));
param.getShelfOrderList().add(shelfOrder);
param.setWarehouseId(asnOrder.getWarehouseId());
param.setType(ShelfOrderTypeEnum.SHELF.getCode());
return param;
\end{lstlisting}

\subsection{Large Language Models}\label{Large Language Models}

To avoid our approach overfitting a certain Large Language Model, we utilize six common LLMs (as shown in Table \ref{tab:LLMs used to experiment}, three open-source and three closed-source models respectively) for code completion. These models operate with default settings designed to enhance result reproducibility: (a) a maximum output token limit of 4096, ensuring extensive output capacity, (b) a temperature setting of zero, which promotes deterministic output by eliminating randomness in response selection, and (c) a \texttt{top\_p} setting of 0.2 configured to focus the model's predictions on the most likely outcomes, thereby improving the precision of the generated code.

\begin{table}[ht]
\centering
\caption{LLMs used to experiment}
\label{tab:LLMs used to experiment}
\begin{tabular}{@{}ccc@{}}
\toprule 
\textbf{Model} & \textbf{Source} & \textbf{Provider}\\ 
\midrule 
GPT-4o & Closed & OpenAI  \\
Gemini-pro-1.5 & Closed & Google  \\
Claude-3.5-haiku & Closed & Anthropic  \\
Llama-3.1-405b & Open & Meta  \\
Qwen-2.5-coder-32b-instruct & Open & Alibaba  \\
Deepseek-3 & Open & Deepseek  \\
\bottomrule 
\end{tabular}
\end{table}

\subsection{Evaluation Metrics}

\subsubsection{CodeBLEU score}

CodeBLEU \cite{ren_codeBLEU_2020} score is an extension of the BLEU score, designed specifically for code completion tasks. In addition to measuring textual similarity, it incorporates Abstract Syntax Tree (AST) and data-flow structures to evaluate both the grammatical correctness and the logical coherence of the generated code, providing a more comprehensive assessment than BLEU alone.

\subsubsection{BLEU score}

BLEU \cite{papineni_BLEU_2002} score is a metric originally developed for evaluating machine translation but has been widely adopted for assessing code completion. In this context, it measures the similarity of n-grams between the generated code and the ground truth. For our evaluation, we use BLEU-4, which compares sequences of four tokens, following the methodology used in previous research.

\subsubsection{Edit Similarity}

Edit Similarity(ES) measures the similarity between two code snippets based on the editing operations. 

\subsubsection{Build Pass}

Build Pass evaluates the correctness of code generated by LLMs due to the limitations of metrics that only compare the reference code and target code snippet in terms of similarity. We employ the compilation mechanism to check if the generated code snippets are runnable at the first stage. If a code script can be compiled, we further use the testing case to invoke the function in each script at the second stage to simply check the output of these functions. A code script will be regarded as successful in Build Pass if the two stages are passed.

\begin{table}[ht]
\centering
\caption{Overall BLEU-4(B4), CodeBLEU(CB), Build Pass(BP) and Edit Similarity(ES) for CCCI and the original prompt}
\label{tab:overall performance}
\begin{tabular}{@{}ccccc@{}}
\toprule 
\textbf{Method} & \textbf{B4(\%)} & \textbf{CB(\%)} & \textbf{ES(\%)} & \textbf{BP(\%)} \\ 
\midrule 
CCCI & \textbf{20.3} & \textbf{41.0} & \textbf{36.7} & \textbf{49.1} \\ 
original & 10.6 & 16.9 & 5.5 & 0.0\\    
\bottomrule 
\end{tabular}
\end{table}

\subsection{RQ1 How effective is the CCCI method when using enhanced prompts tailored with contextual information?}\label{Effectiveness Evaluation}

We hypothesized that the CCCI method, by utilizing enhanced prompts enriched with contextual information, would significantly improve the effectiveness of code completion as measured by BLEU-4, CodeBLEU, Edit Similarity and Build Pass scores. This enhancement is expected to lead to a better alignment with the real software development tasks, thus producing more accurate and functional code than the original prompt without class information retrieved from the current project and its dependencies. To test this hypothesis, we selected 289 production code scripts and used the CCCI method to regenerate corresponding scripts. The effectiveness of the regenerated scripts was quantified by calculating the average BLEU-4, CodeBLEU, Edit Similarity and Build Pass scores across these samples.

\subsubsection{BLEU-4 Score Result}

The average BLEU-4 score for CCCI-generated scripts is 20.3, which is substantially higher than the original prompt's average of 10.6. This 91.5\% increase in score highlights CCCI's enhanced capability in accurate code completion through n-gram comparison.

\subsubsection{CodeBLEU Result}

Similarly, the average CodeBLEU score for CCCI is 41.0, compared to 16.9 for the original prompt, reflecting an improvement of 142.6\%. This improvement indicates more grammatical correctness and logic correctness within the generated scripts, demonstrating CCCI's superior contextual integration.

\subsubsection{Edit Similarity Result}

The average Edit Similarity has improved significantly from 5.5 to 36.7. The noticeable improvement indicates that developers can use the code scripts with less effort for code modification based on the generated code scripts to satisfy their requirements. 

\subsubsection{Build Pass Result}

As Table \ref{tab:overall performance} shows, the code scripts generated through the original prompt cannot be either compiled or tested, which demonstrates the challenges for code completion without the contextual information in real industrial settings.

The higher average scores for a series of metrics above validate CCCI's effectiveness in integrating contextual information retrieved from the current project into code completion. The variability in CodeBLEU scores is predominantly attributed to a main factor observed in the original scripts:

Custom Code Exception Handling: Many scripts utilized custom exceptions for error handling (e.g., Listing 7), which were not adequately captured by CCCI. This discrepancy arose because the contextual information provided did not include sufficient details on the custom exception-handling classes and methods. Enhancing the contextual richness to include these specifics could potentially improve performance in scenarios involving custom error handling.

\textbf{Listing 7: An example of a script with custom code exception handling}
\begin{lstlisting}[language=Java, linewidth=\columnwidth]
if (CollectionUtils.isEmpty(inboundOrderList)) {
    throw INVENTORY_NOT_FOUND.instanceException();
}
InboundOrder inboundOrder = inboundOrderList.get(0);
if (!InboundOrderStatusEnum.DEFECTIVE_WAIT_CONFIRMED
.getCode().equals(inboundOrder.getStatus())) {
    throw INVENTORY_STATUS_ERROR.instanceException();
}
\end{lstlisting}

\begin{table}[H]
\centering
\caption{BLEU-4(\%) for CCCI with Different LLMs}
\label{tab:bleu-4_comparison}
\begin{tabular}{@{}cccc@{}}
\toprule 
\textbf{Model} & \textbf{Original} & \textbf{CCCI} & \textbf{$\uparrow$
 (\%)} \\  
\midrule 
GPT-4o & 10.6 & 20.3 & 91.5 \\  
Gemini-pro-1.5 & 6.2 & 22.1 & 256.5 \\ 
Claude-3.5-haiku & 7.4 & 21.1 & 185.1 \\ 
Llama-3.1-405b & 14.2 & 25.0 & 76.1 \\  
Qwen-2.5-coder-32b & 14.0 & 22.7 & 62.1 \\ 
Deepseek-3 & 11.4 & 23.6 & 107.0 \\  
\bottomrule 
\end{tabular}
\end{table}

\begin{table}[H]
\centering
\caption{CodeBLEU(\%) for CCCI with different LLMs}
\label{tab:CodeBLEU_comparison}
\begin{tabular}{@{}cccc@{}}
\toprule 
\textbf{Model} & \textbf{Original} & \textbf{CCCI} & \textbf{$\uparrow$
 (\%)} \\  
\midrule 
GPT-4o & 16.9 & 41.0 & 142.6 \\  
Gemini-pro-1.5 & 28.8 & 38.0 & 31.9 \\  
Claude-3.5-haiku & 19.1 & 34.5 & 80.6 \\  
Llama-3.1-405b & 18.9 & 41.0 & 116.9 \\  
Qwen-2.5-coder-32b & 21.3 & 40.4 & 89.7 \\  
Deepseek-3 & 24.2 & 41.7 & 72.3 \\  
\bottomrule 
\end{tabular}
\end{table}

\begin{table}[H]
\centering
\caption{Edit Similarity(\%) for CCCI with different LLMs}
\label{tab:editsimilarity_comparison}
\begin{tabular}{@{}cccc@{}}
\toprule 
\textbf{Model} & \textbf{Original} & \textbf{CCCI} & \textbf{$\uparrow$
 (\%)} \\  
\midrule 
GPT-4o & 5.5 & 36.7 & 567.3 \\  
Gemini-pro-1.5 & 4.5 & 36.9 & 720.0 \\ 
Claude-3.5-haiku & 4.6 & 24.0 & 421.7 \\ 
Llama-3.1-405b & 5.2 & 42.7 & 721.2 \\  
Qwen-2.5-coder-32b & 5.4 & 38.4 & 611.1 \\ 
Deepseek-3 & 5.5 & 38.2 & 594.5 \\  
\bottomrule 
\end{tabular}
\end{table}

\begin{table}[H]
\centering
\caption{Build Pass Rate(\%) for CCCI with different LLMs, there is no single script produced by the original prompt passing the Build Pass test}
\label{tab:buildingpass_comparison}
\begin{tabular}{@{}cccc@{}}
\toprule 
\textbf{Model} & \textbf{Total} & \textbf{Build Pass} & \textbf{Pass Rate (\%)} \\  
\midrule 
GPT-4o & 289 & 142 & 49.1 \\  
Gemini-pro-1.5 & 289 & 185 & 64.0 \\  
Claude-3.5-haiku & 289 & 75 & 26.0 \\ 
Llama-3.1-405b & 289 & 150 & 51.9 \\  
Qwen-2.5-coder-32b & 289 & 99 & 34.3 \\  
Deepseek-3 & 289 & 83 & 28.7 \\  
\bottomrule 
\end{tabular}
\end{table}

\subsection{4.4 RQ2 What is the impact of the CCCI method across different large-scale language models?}\label{Impact Evaluation}

The motivation behind RQ2 is to demonstrate the adaptability of the CCCI method across different large-scale language models (LLMs \cite{openai_llm_2023}), as validated on six popular open-source or closed-source models: ChatGPT4 \cite{openai_chatgpt_2022, openai_api_2023}, Gemini-pro \cite{google_gemini_2023}, Claude-3 \cite{anthropic_claude3_2023}, Deepseek-3 \cite{bi2024deepseek}, Llama-3.1 \cite{touvron_llama3_2024}, and Qwen-2.5 \cite{qwen2.5}. By evaluating the CodeBLEU, BLEU-4, Edit Similarity, and Build Pass scores, we aim to prove that CCCI can consistently generate accurate and relevant code across a variety of LLMs, highlighting its robustness in different model environments.
As shown in the tables above, most models achieve similar scores, with CodeBLEU, BLEU-4, Edit Similarity, and Build Pass results relatively close across models such as GPT-4o, Gemini-pro, Llama-3.1, Deepseek-3, Qwen-2.5. However, Claude-3.5-haiku significantly underperforms, with lower scores in these metrics. This discrepancy is attributed to Claude-3.5-haiku's weaker instruction-following capabilities, while the other models demonstrate comparable performance, reinforcing the CCCI method’s effectiveness across various LLMs.

\section{Related Work}\label{Related Work}

The field of code completion using large language models (LLMs) has seen significant advancements in recent years. This section reviews relevant literature that contributes to understanding the current state of research in this domain.
Various approaches have been explored for code completion. Sequence-based methods generate code token by token based on input descriptions, while tree-based methods construct parse trees from natural language descriptions and convert them into code. Recent models like CodeT5 and CodeGPT leverage the transformer architecture to enhance the quality of generated code \cite{svyatkovskiy_compose_2020, sun_treegen_2020}.
Pre-trained language models such as CodeBERT, CodeT5, InCoder, and CodeGPT have been pivotal in advancing code generation tasks. These models are typically fine-tuned for specific tasks, such as generating code from natural language descriptions, completing code snippets, and generating unit tests. For instance, CodeBERT and CodeT5 have shown substantial improvements in generating accurate code by leveraging large-scale pre-training on code and natural language data.
Retrieval-augmented language models \cite{chen_code_2024, hayati_retrieval-based_2018, lu_reacc_2022, parvez_retrieval_2021} on for their ability to improve the performance of code completion tasks by incorporating relevant external information. For example, the REDCODER framework retrieves relevant code snippets or summaries to enhance the performance of code completion models. This approach has been shown to improve the accuracy and relevance of generated code by providing additional context to the model.

Recent research has also focused on domain-specific code completion, such as generating unit tests and library-oriented code. For instance, the study on using large language models for automated unit test generation demonstrates the potential of these models to significantly reduce the effort required for test creation by generating high-quality unit tests automatically. Similarly, the CodeGen4Libs approach presents a two-stage method for generating code that interacts with third-party libraries, addressing the challenges of library-specific code completion.

\section{Threats to Validity}\label{Threats to Validity}

\textbf{Quality of the Dataset.} The dataset used in this study comprises code that is currently deployed in production within a real industrial sector; we extract these code scripts from the WMS project. While this ensures that our findings are grounded in practical, real-world applications, it also poses a limitation: the results obtained might not be generalizable to other datasets or environments. The specific characteristics and challenges of the industrial dataset may influence the performance metrics, and thus, transferring our approach to a different context might yield different outcomes.

\textbf{Parameter Specification in Models.} In our experiments, the top\_p and temperature parameters were set to 0.2 and zero based on preliminary tests designed to stabilize results and mitigate the risk of variability. However, this parameter setting may not be optimal for other datasets or different types of tasks. While the selection of 289 scripts was intended to provide a robust average by minimizing the effects of outliers, different configurations or datasets might require adjustments to this parameter to achieve the best results. In addition, we use all the scripts with lengths between 300 and 700 because of the limitation of the LLMs(the shorter length will be just lines of code, which makes the metrics unreliably high; the longer length will cause the LLMs generating code with fewer details), and the invocation speed also poses a challenge to experiment with all the scripts. For example, we invoke LLMs 12 times for every script; compiling and testing the script is also time-consuming. In general, the whole experiment takes more than 8 hours for these 289 scripts. 

\section{Conclusion}\label{Conclusion}

This research introduced the CCCI method, a novel approach aimed at enhancing the quality of automated code completion. By retrieving 289 scripts from over 819 operational scripts currently deployed across internet enterprise applications, we demonstrated a notable improvement over many large language models in CodeBLEU, BLEU-4, Edit Similarity, and Build Pass scores. These results emphasize the practical applicability and effectiveness of our approach in real-world settings. 

Furthermore, our evaluation across multiple large-scale language models demonstrated the adaptability of the CCCI method, as it consistently generated accurate code with similar performance across most models, further validating its robustness and versatility in different LLM environments.

We do not compare our approach to the prior studies such as RLPG \cite{repository_prompt_generation_2023} because they focus on different dimensions of code completion. For example, RLPG completes line-level code in the current repository, and our method completes body-level code scripts that are more complex than line-level code completion. Comparing with popular retrieval-based approaches is not suitable because our dataset is not generic enough to compare. In addition, we conduct the evaluation on various metrics, and we employ different LLMs to evaluate our approach, which is commonly absent in prior approaches. 

Despite being tailored primarily for data mapping tasks, the principles underlying the CCCI method apply to broader scenarios, such as Service Mashup. This versatility presents a potential path for future research, exploring the adaptation and application of CCCI in various other contexts where automated, context-aware code completion can play a pivotal role.

\printbibliography

\end{document}